\documentclass[twocolumn,showpacs,preprintnumbers,amsmath,amssymb,prl]{revtex4-1}
\usepackage{dcolumn}
\usepackage{upgreek} 
\usepackage{txfonts} 
\usepackage{ifpdf}
\ifpdf
   \usepackage[pdftex]{graphicx}
  \else
    \usepackage[dvips]{graphicx}
\fi 
\usepackage{enumerate}

\begin{document}

\title{Temperature dependence of geometrical and velocity matching resonances in Bi$_2$Sr$_2$CaCu$_2$O$_{8+x}$ intrinsic Josephson junctions}

\author{S. O. Katterwe}
\author{V. M. Krasnov}

\affiliation{Department of Physics, Stockholm University, AlbaNova University Center, SE-10691 Stockholm, Sweden}

\date{\today}

\begin{abstract}
We study temperature dependence of geometrical (Fiske) and
velocity-matching (Eck) resonances in the flux-flow state of small
$\mathrm{Bi_2Sr_2CaCu_2O_{8+x}}$ mesa structures. It is shown that
the quality factor of resonances is high at low $T$, but rapidly
decreases with increasing temperature already at $T > 10$ K. We
also study $T$-dependencies of resonant voltages and the speed of
electromagnetic waves (the Swihart velocity). Surprisingly it is
observed that the Swihart velocity exhibits a flat $T$-dependence
at low $T$, following $T-$dependence of the $c$-axis critical
current, rather than the expected linear $T$-dependence of the
London penetration depth. Our data indicate that self-heating is
detrimental for operation of mesas as coherent THz oscillators
because it limits the emission power via suppression of the
quality factor. On the other hand, significant temperature
dependence of the Swihart velocity allows broad-range tunability
of the output frequency.

\end{abstract}


\maketitle

\section{I. Introduction}

Single crystals of a cuprate superconductor
$\mathrm{Bi_2Sr_2CaCu_2O_{8+x}}$ (Bi-2212) represent natural
stacks of atomic scale intrinsic Josephson junctions (IJJs)
\cite{Kleiner92}. Josephson junctions form transmission lines for
electromagnetic (EM) waves \cite{Swihart}. The propagation
(Swihart) velocity is $c_0 \simeq c/[L_{\square}
C_{\square}]^{1/2}$, where $c$ is the speed of light in vacuum and
$L_{\square}$ and $C_{\square}$ are the inductance and the
capacitance per square of the transmission line.
\begin{eqnarray}\label{Lsquare}
L_{\square}= 4\pi \Lambda, \\
\Lambda= t + 2\lambda_S \coth (d/\lambda_S).
\end{eqnarray}
Here $t$ and $d$ are thicknesses of the dielectric and
superconducting layers, respectively, and $\lambda_S$ is the
London penetration depth of the superconductor. In thin-layer
junctions $t,d \ll \lambda_S$, $L_{\square} \simeq
8\pi\lambda_S^2/d$ is dominated by a large kinetic inductance of
superconducting layers. As a consequence, $c_0$ can be much slower
than $c$ - the phenomenon that finds applications in compact
superconducting delay lines \cite{DelayLine}.

The Swihart velocity carries a direct information about the London
penetration depth. It can be obtained by measuring the propagation
(delay) time in a transmission line \cite{Mason,Kircher}. However,
Josephson junctions provide a much easier way of measuring $c_0$.
In Josephson junctions EM waves can be generated in-situ by means
of the ac-Josephson effect. At geometrical resonance conditions
they form standing waves, leading to appearance of Fiske steps in
current-voltage ($I$-$V$) characteristics
\cite{Dmitrenko,Koshelets,Cirillo,FiskeInd,KatterweFiske}. Fiske
step voltages allow simple and direct evaluation of the {\it
absolute values} of $\lambda_S(T)$ \cite{Dmitrenko,Ngai} (unlike
surface impedance measurements, which usually provide only
relative values \cite{Lambda,Lambda2,Trunin,Lambda2005}).
Such measurements do not require long transmission lines, but can
be performed on small $\sim\mu$m-scale junctions. IJJ of $\mu$m
sizes, made on high quality Bi-2212 single crystals, are free from
crystallographic defects, that can affect $\lambda$ in cuprates
\cite{Pan,Prozorov}. Therefore, Fiske resonances in small IJJs
should provide information about genuine (defect-free) behavior of
the penetration depth in cuprates.

Geometrical resonances play also an important role in achieving
high power THz EM wave emission from Bi-2212 mesa structures
\cite{Ozyuzer,Wang,Nonequilibrium,Hu,Klemm,Breather}. The maximum
radiation power from a stack with $N$ junctions is $P_{rad}
\propto N^2 Q^2$ \cite{TheoryFiske}, where
\begin{equation}
Q=\omega RC, \label{eq:Q}
\end{equation}
is the quality factor of the resonance, $\omega$ is the resonant
frequency, $R$ the effective damping resistance and $C$ the
capacitance of the junctions. The factor $N^2$ is due to
constructive interference of $N$ in-phase synchronized junctions
\cite{Note1} and the factor $Q^2$ represents the resonant
amplification in each junction by the geometrical resonance. Thus,
both the in-phase coherence and the high quality $Q\gg 1$
geometrical resonances are needed for achieving high emission
power \cite{TheoryFiske}. Increment of the emission power is
inevitably accompanied by self-heating of the stack. In
superconductors this leads to a rapid increment of the
quasiparticle (QP) damping, which suppresses $Q$. Self-heating
ultimately limits the performance of an oscillator
\cite{Breather}. Clearly, investigation of the quality factor of
geometrical resonances and their $T$-dependence has a primary
significance for development of high power THz oscillator, based
on IJJs.

In this work, we study experimentally $T$-dependencies of
geometrical (Fiske) and velocity-matching (Eck) resonances
\cite{Cirillo} in the flux-flow state of small Bi-2212 mesa
structures. It is observed that $Q$ of resonances is large at low
$T$, but rapidly decreases with increasing temperature already at
$T \gtrsim 10$ K $\ll T_c \sim 90$ K, primarily due to enhancement
of the quasiparticle damping. Surprisingly, it is observed that
resonant voltages, proportional to the Swihart velocity, exhibit a
very weak $T$-dependence at low $T$ and do not follow the expected
linear $T$-dependence of the effective London penetration depth
$\lambda_{ab}(T)$ in Bi-2212
\cite{Lambda,Lambda2,Trunin,Lambda2005}. We discuss possible
origins of such a distinct discrepancy, which to our opinion
deserves further experimental and theoretical analysis.

\section{II. Geometrical resonances in stacked Josephson junctions}

Stacked Josephson junctions form multilayer transmission lines for
electromagnetic waves. The general problem of linear wave
propagation in multilayer transmission lines was first considered
by Economou \cite{Economou} and more recently within the
inductively coupled junction (ICJ) formalism by Kleiner
\cite{KleinerModes} and Sakai et al., \cite{SakUstFiske}.
In this section we will briefly recollect peculiarities of wave
propagation and geometrical resonances in stacked Josephson
junctions.

In the ICJ model of Sakai, Bodin and Pedersen \cite{SBP}, a
layered superconductor is represented by a stack of isotropic
superconducting layers with the thickness $d$ and the
``intrinsic'' penetration depth $\lambda_S$, separated by tunnel
barriers with the thickness $t$, the dielectric constant
$\epsilon_r$, and the fluctuation-free (maximum) Josephson
critical current density $J_{c0}$. The stacking periodicity
$s=t+d$ is $\simeq 1.5$ nm for Bi-2212. Properties of inductively
coupled stacked Josephson junctions are described by the coupled
sine-Gordon equation \cite{SBP}. The coupling is represented by a
tridiagonal coupling matrix $A$ with the of-diagonal terms equal
to minus the effective inductive coupling constant between
neighbor junctions \cite{SakUstFiske},
\begin{equation}\label{CouplingConst}
S= \lambda_S \left[ t\sinh\left(
\frac{d}{\lambda_S}\right)+2\lambda_S\cosh\left(
\frac{d}{\lambda_S}\right)\right]^{-1}.
\end{equation}
For atomic scale IJJs, $S\simeq 0.5-ds/4\lambda_S^2$ is very close
to its maximum value 0.5.

\subsection{A. Eigen-modes in stacked junctions}

The main difference between single and stacked junctions is the
presence of multiple electromagnetic wave modes in the stack.
Geometrical resonances in a stack correspond to formation of
two-dimensional standing waves \cite{KleinerModes,SakUstFiske}.
The wave number along the $ab$-planes ($x$-axis) is $k_m=\pi m
/L$, where $L$ is the length of the junctions and $m$ is the
number of nodes in the standing wave. In the $c$-axis direction it
is given by one of the eigen-modes, $k_n=\pi n/(N+1)s$,
$n=1,2,...N$, where $N$ is the number of junctions in the stack.
The oscillatory part of the phase difference is:
\begin{equation}\label{GeomModes}
\delta \varphi_{i}(m,n) = a \cos\left(\frac{\pi m x}{L} \right)
\sin\left(\frac{\pi n i}{N+1} \right)e^{j\omega t}.
\end{equation}
Here $i=1,2,...N$ is the junction index, $a=$const is an
amplitude, and $\omega$ is the angular frequency.

Each eigen-mode has a distinct propagation velocity, given by Eq.
(3.52) of Ref. \cite{Economou}. Within the ICJ model they can be
written as \cite{SakUstFiske}:
\begin{equation}\label{c_n}
c_n=c_0\left[1-2S\cos\left( \frac{\pi n}{N+1}\right)
\right]^{-1/2}, ~n=1,2,...N,
\end{equation}
The Swihart velocity $c_0= \lambda_J \omega_p$ where
\begin{equation}\label{OmegaP}
\omega_p=\left[\frac{8\pi^2 t c J_c}{\Phi_0
\epsilon_r}\right]^{1/2},
\end{equation}
is the Josephson plasma frequency and
\begin{equation}\label{Lambda_J}
\lambda_J=\left[\frac{\Phi_0c}{8\pi^2 J_{c0}\Lambda}
\right]^{1/2}\simeq\left[\frac{\Phi_0cs}{16\pi^2
J_{c0}\lambda_{ab}^2} \right]^{1/2}
\end{equation}
is the Josephson penetration depth of a single junction and
\begin{equation}\label{lambda_ab}
\lambda_{ab}\simeq \lambda_S\sqrt{s/d}
\end{equation}
is the effective London penetration depth for field perpendicular
to layers.

Similarly, eigen-modes are characterized by different
characteristic lengths \cite{fluxon}
\begin{equation}\label{LambdaN}
\lambda_n=\lambda_J\left[1-2S\cos\left( \frac{\pi m}{N+1}\right)
\right]^{-1/2}, ~n=1,2,...N.
\end{equation}
$(\lambda_J/\lambda_n)^{2}$ are eigenvalues of the coupling matrix
$A$ \cite{fluxon}.
The shortest, $\lambda_N \simeq \lambda_J/\sqrt{2} \simeq 0.5~
\mu$m for Bi-2212. The longest $\lambda_1$
approaches the effective penetration depth for field parallel to
layers
\begin{equation}\label{Lambda_c}
\lambda_c = \left[\frac{\Phi_0c}{8\pi^2 J_{c0} s} \right]^{1/2},
\end{equation}
for $N\gg \pi\lambda_{ab}/s \simeq 400$. In Bi-2212,
$\lambda_c(T=0)\sim 100~ \mu$m $\gg \lambda_{ab}(T=0)\simeq
0.2~\mu$m \cite{fluxon}.

\begin{figure}[t]
\includegraphics[width=0.48\textwidth]{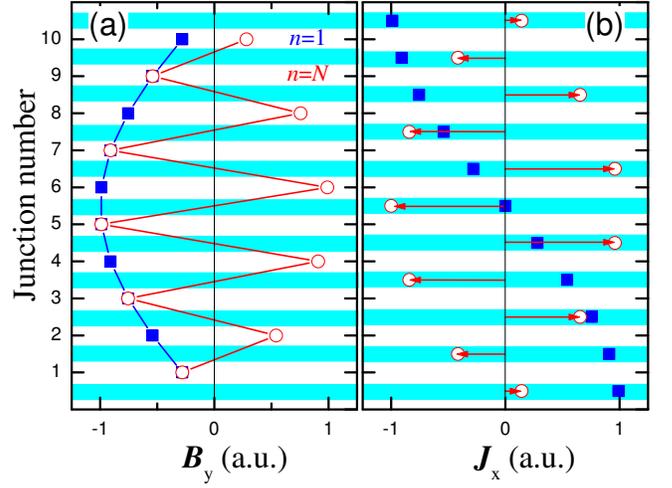}
\caption{\label{fig1} (Color online). Spatial distribution of
oscillation amplitudes of (a) magnetic field and (b) in-plane
currents for the in-phase (squares) and the out-of-phase (circles)
modes for a stack with $N=10$ IJJs. Horizontal stripes represent
superconducting layers.}
\end{figure}

Due to inductive coupling between junctions, the in-plane
($y$-axis) magnetic field is non-local and depends on phase
distributions in all junctions: $B_y(i) = (H_0/2)A^{-1}\lambda_J
\partial\varphi_j / \partial x$. Here $H_0=\Phi_0/\pi \lambda_J
\Lambda$ \cite{fluxon}. Using Eq.(\ref{GeomModes}) we obtain for
the oscillatory part of magnetic field in the stack:
\begin{equation}\label{By(i)}
B_y(x,z)(m,n) = -\frac{H_0 a \pi m \lambda_n^2}{2L\lambda_J}
\sin\left(k_m x \right) \sin\left(k_n z\right).
\end{equation}
Here we used the property that $A$ and $A^{-1}$ have the same
eigenvectors, and eigenvalues of $A^{-1}$ are
$\lambda_n^2/\lambda_J^2$, Eq. (\ref{LambdaN}).

The in-plane current density in superconducting layers is obtained
from the Maxwell equation $J_x=-(c/4\pi)
\partial B_y/\partial z$:
\begin{eqnarray}\label{Jx(i)}
J_x(x,z)(m,n) = J_{ac}(m,n)\sin\left(k_m x
\right) \cos\left(k_n z\right), \\
J_{ac}(m,n)=\frac{a\Phi_0 c \lambda_n^2 m n}{16 \lambda_{ab}^2
\lambda_J L (N+1)}.
\end{eqnarray}

Fig. \ref{fig1} shows calculated distributions of the amplitudes
of $B_y$ (a) and $J_x$ (b) for modes $n=N$ (open circles) and
$n=1$ (squares) for the stack with $N=10$ junctions. Horizontal
stripes represent superconducting layers. It is seen that the
eigen-modes are characterized by different symmetry along the
stacking direction. The slowest $n=N$ mode corresponds to the
(almost) out-of-phase state in neighbor junctions $\delta
\varphi_{i} \simeq -\delta \varphi_{i+1}$ . The fastest $n=1$ mode
corresponds to the (almost) in-phase state $\delta \varphi_{i}
\simeq \delta \varphi_{i+1}$.

\begin{figure}[t]
\includegraphics[width=0.4\textwidth]{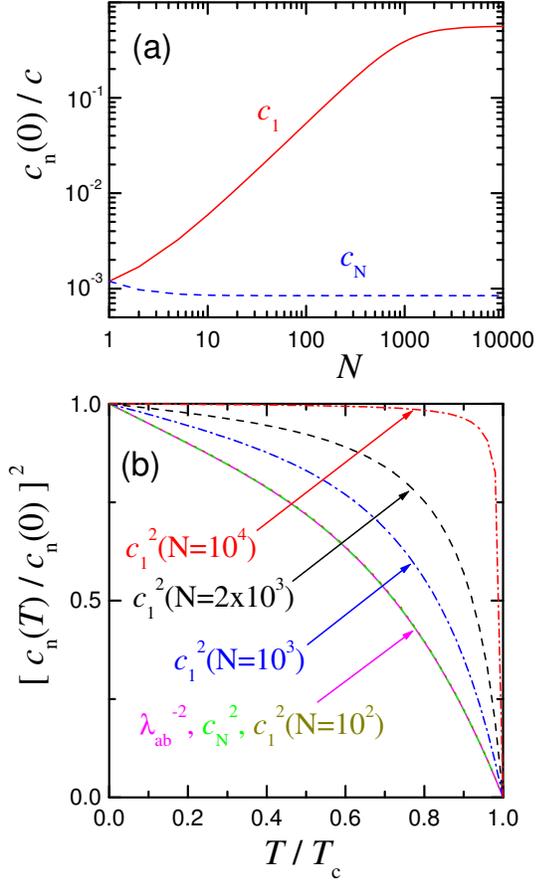}
\caption{\label{fig2} (Color online). (a) Calculated slowest and
fastest velocities $c_N$ and $c_1$ as a function of the number of
junctions in the stack. Calculations are made within the ICJ model
for typical Bi-2212 parameters at $T=0$. (b) Calculated
temperature dependencies of squares of $c_N$ and $c_1$ for several
$N$ along with $\lambda_{ab}^{-2}(T)$, normalized by the
corresponding values at $T=0$. }
\end{figure}

Fig. \ref{fig2} (a) shows calculated dependence of $c_1$ and $c_N$
on the number of junctions $N$. It is seen that the slowest
velocity is almost independent of $N$ \cite{KatterweFiske}
\begin{equation}\label{cN}
c_N \simeq \frac{c_0}{\sqrt{2}}\simeq c
\left[\frac{ts}{4\varepsilon_r \lambda_{ab}^2 }\right]^{1/2}.
\end{equation}
 To the contrary, the fastest velocity
\begin{equation}\label{c1}
c_1 \simeq c \sqrt{\frac{t}{\varepsilon_r s}}
\left[1+\left(\frac{\pi\lambda_{ab}}{s(N+1)}\right)^2\right]^{-1/2}
\end{equation}
is growing linearly with $N$ for $N < \pi\lambda_{ab}/s \simeq
400$ \cite{KatterweFiske}. 
For $N \gg \pi\lambda_{ab}(T)/s$, it asymptotically approaches the
$T$-independent value $c_1(N\rightarrow \infty) =
c[t/\varepsilon_r s]^{1/2}$, close to the speed of light in the
dielectric, as shown in Fig. \ref{fig2} (a).

Fig. \ref{fig2} (b) shows calculated $T$-dependencies of $c_1^2$
and $c_N^2$, Eq. (\ref{c_n}), normalized on the corresponding
values at $T=0$, for different $N$. Calculations are made for
typical parameters of Bi-2212, using the
$[\lambda_{ab}(T)/\lambda_{ab}(0)]^{-2}$ dependence shown by the
lowest line obtained from surface impedance measurements
\cite{Lambda,Lambda2,Trunin,Lambda2005,Prozorov}. As follows from
Eq. (\ref{cN}), $T$-dependence of the out-of-phase velocity $c_N$
follows $1/\lambda_{ab}(T)$, irrespective of $N$. For IJJs the
same is true for all slow modes $n\geqslant 2$.

The speed of the fastest mode, $c_1(T)$, does depends on $N$. For
$N < \pi\lambda_{ab}/s \simeq 400$, it maintains the same
$T$-dependence $\propto 1/\lambda_{ab}$. The corresponding three
curves $[c_N(T)/c_N(0)]^2$,
$[\lambda_{ab}(T)/\lambda_{ab}(0)]^{-2}$ and $[c_1(T)/c_1(0)]^2$
for $N=100$ collapse in one in Fig. \ref{fig2} (b). For much
larger $N$, when $c_1$ approaches $T$-independent speed of light
in the dielectrics, see Fig. \ref{fig2} (a), $c_1(T)$ becomes
flatter at low $T$, as shown in Fig. \ref{fig2} (b). However,
since $\lambda_{ab}$ diverges at $T\rightarrow T_c$, $c_1$ always
vanishes at $T_c$, as seen from the curve with $N=10^4$ in Fig.
\ref{fig2} (a).

In applied in-plane magnetic field Josephson vortices (fluxons)
\cite{fluxon} enter into the junctions. In strong enough magnetic
field fluxons form a regular fluxon lattice in a stack. Usually a
triangular lattice is most stable due to fluxon repulsion. However
a rectangular lattice can be stabilized via geometrical
confinement is small Bi-2212 mesas \cite{Katterwe}. Motion of
fluxons leads to appearance of the flux-flow (FF) branch in the
$I$-$V$. Emission of EM waves in the FF state leads to excitation
of geometrical resonances \cite{Koshelets,Cirillo,KatterweFiske}.
The corresponding Fiske step voltage for the resonant mode $(m,n)$
is
\begin{equation}
V_{m,n}(T)=\Phi_0 m c_n(T) /2L. \label{eq:Vmn}
\end{equation}

The strongest resonance occurs at the velocity matching (VM)
condition, when the velocity of fluxons is equal to the velocity
of electromagnetic waves \cite{KatterweFiske}. This leads to
appearance of the VM (Eck) step at the end of FF branch
\cite{Cirillo}. The VM voltage is
\begin{equation}\label{VMstep}
V_{VM} \simeq N H s c_n.
\end{equation}
The $T$-dependencies of both Fiske and VM steps are determined
solely by $c_n(T)$, Eq.(\ref{c_n}). Therefore they can be used for
accurate detection of the absolute value of $\lambda_{ab}(T)$
(except for the fastest mode at very large $N$, as shown in Fig.
\ref{fig2} (b)).

\subsection{B. Connection between the inductively coupled and the Lawrence-Doniach models}

A similar system of coupled sine-Gordon equations was also
obtained from the Lawrence-Doniach (LD) model \cite{LD}. The two
main parameters of the LD model are the anisotropy factor $\gamma
=$const$\gg 1$ and the effective London penetration depths
$\lambda_{ab}$. The rest of parameters are derived as \cite{LD}:
$\lambda_c=\gamma \lambda_{ab}$, $\lambda_J=\gamma s$,
$\omega_p=c/\varepsilon_r^{1/2}\gamma\lambda_{ab}$ and
$c_0=cs/\varepsilon_r^{1/2}\lambda_{ab}$.

From comparison with ICJ expressions Eqs.
(\ref{OmegaP},\ref{Lambda_J},\ref{lambda_ab},\ref{Lambda_c},\ref{cN})
it is seen that while the ICJ model contains two $T-$dependent
variables $\lambda_{ab}(T)$ and $J_{c0}(T)$, the LD model has only
one, $\lambda_{ab}(T)$, which imposes its $T$-dependence on all
other variables. Within the range of validity of the LD model,
$T_c - T\ll T_c$, the two models are identical because
$\lambda_{ab}^{-2}(T)\propto J_{c0}(T) \propto 1-T/T_c$. However,
as will be discussed below, $\lambda_{ab}^{-2}(T)$ and $J_{c0}(T)$
have distinctly different $T$-dependencies at low $T$, which does
cause a discrepancy between the two models. Essentially it is
related to the fact that in cuprates the anisotropy $\gamma(T)=
\lambda_c(T)/\lambda_{ab}(T) \neq$const
\cite{Lambda_abc,Lambda_c,Radtke}.

\begin{figure}[b]
\begin{center}
   \includegraphics[width=0.23\textwidth]{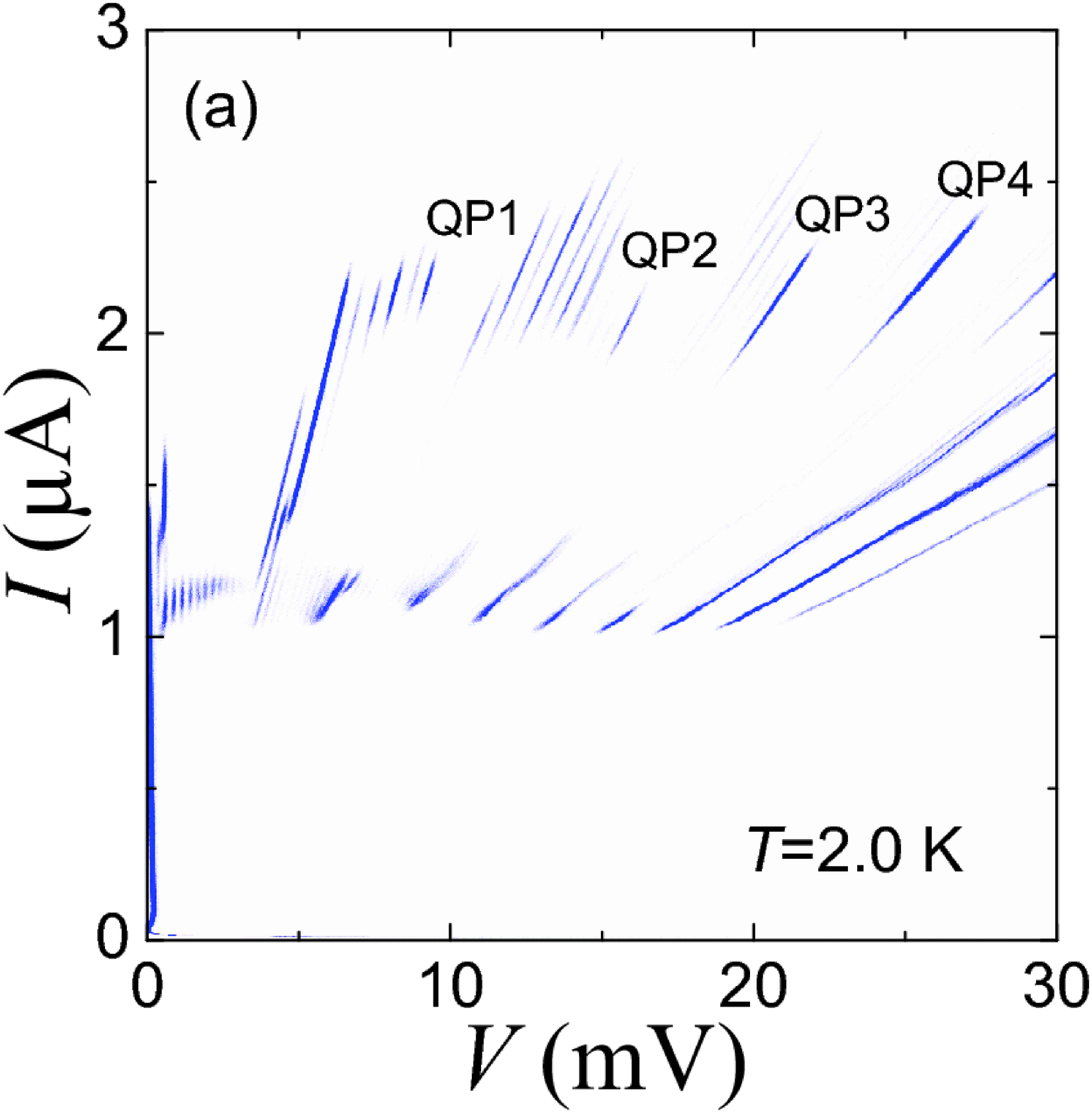}
   \includegraphics[width=0.23\textwidth]{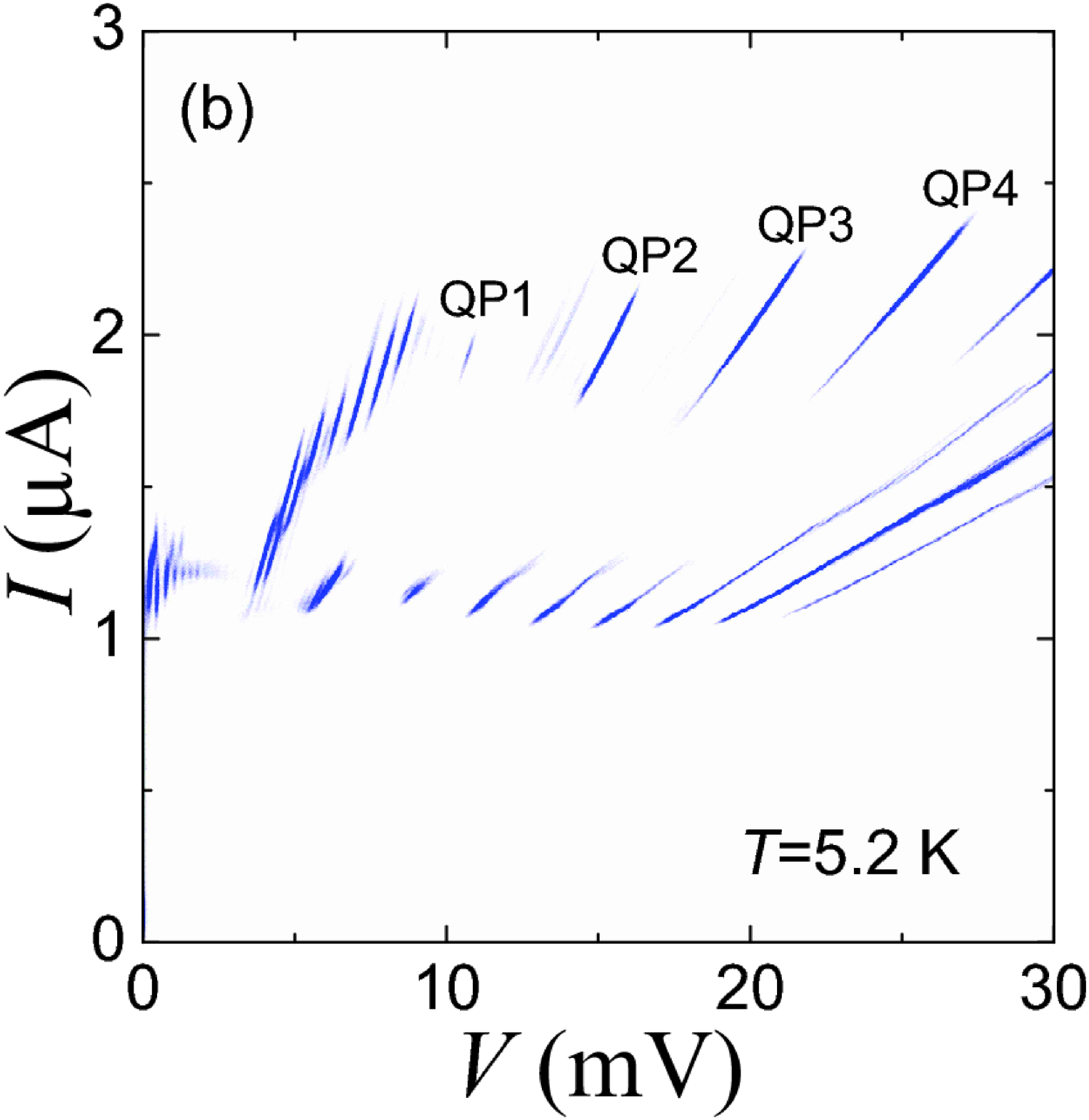}\\
   \includegraphics[width=0.23\textwidth]{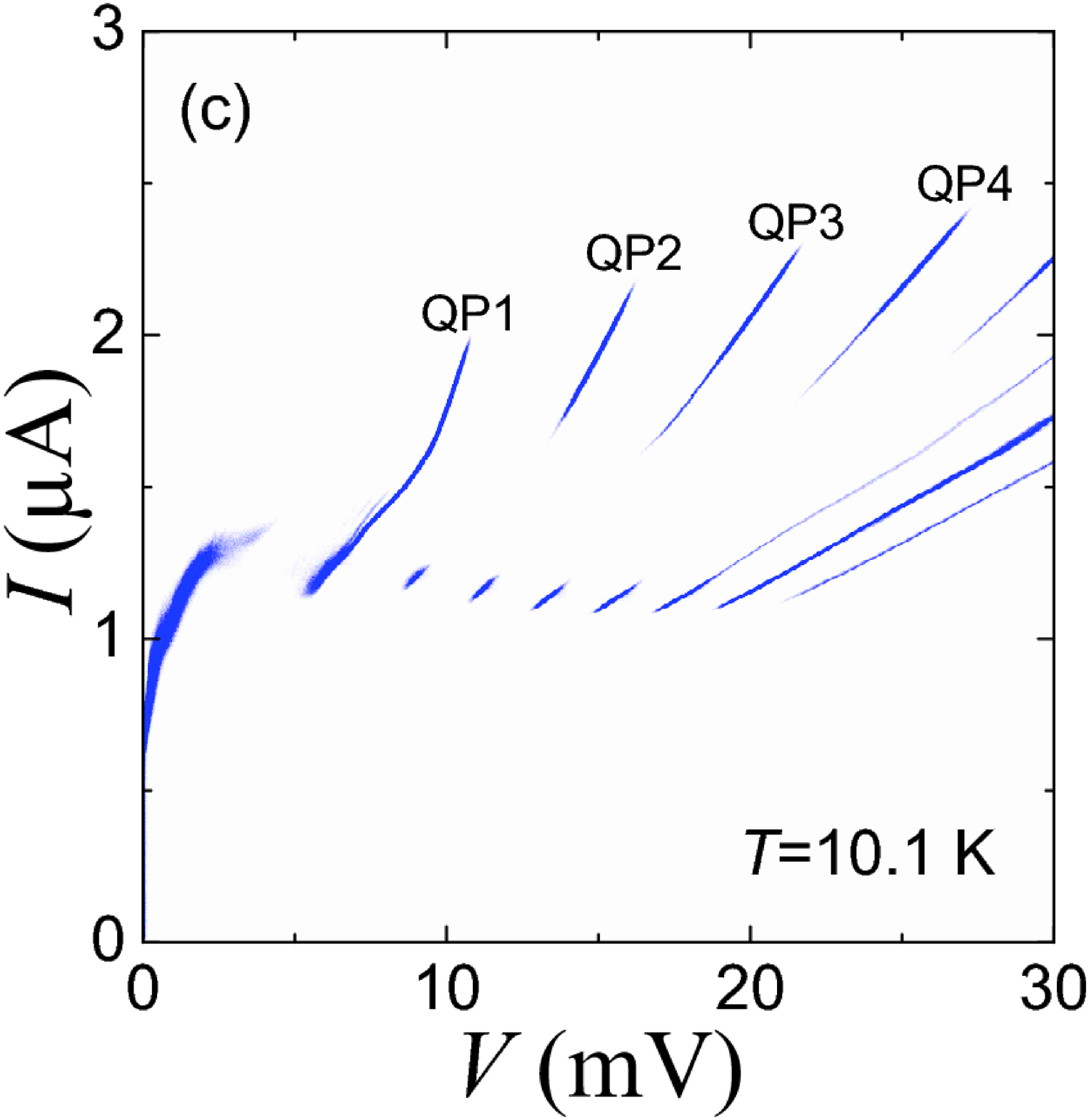}
   \includegraphics[width=0.23\textwidth]{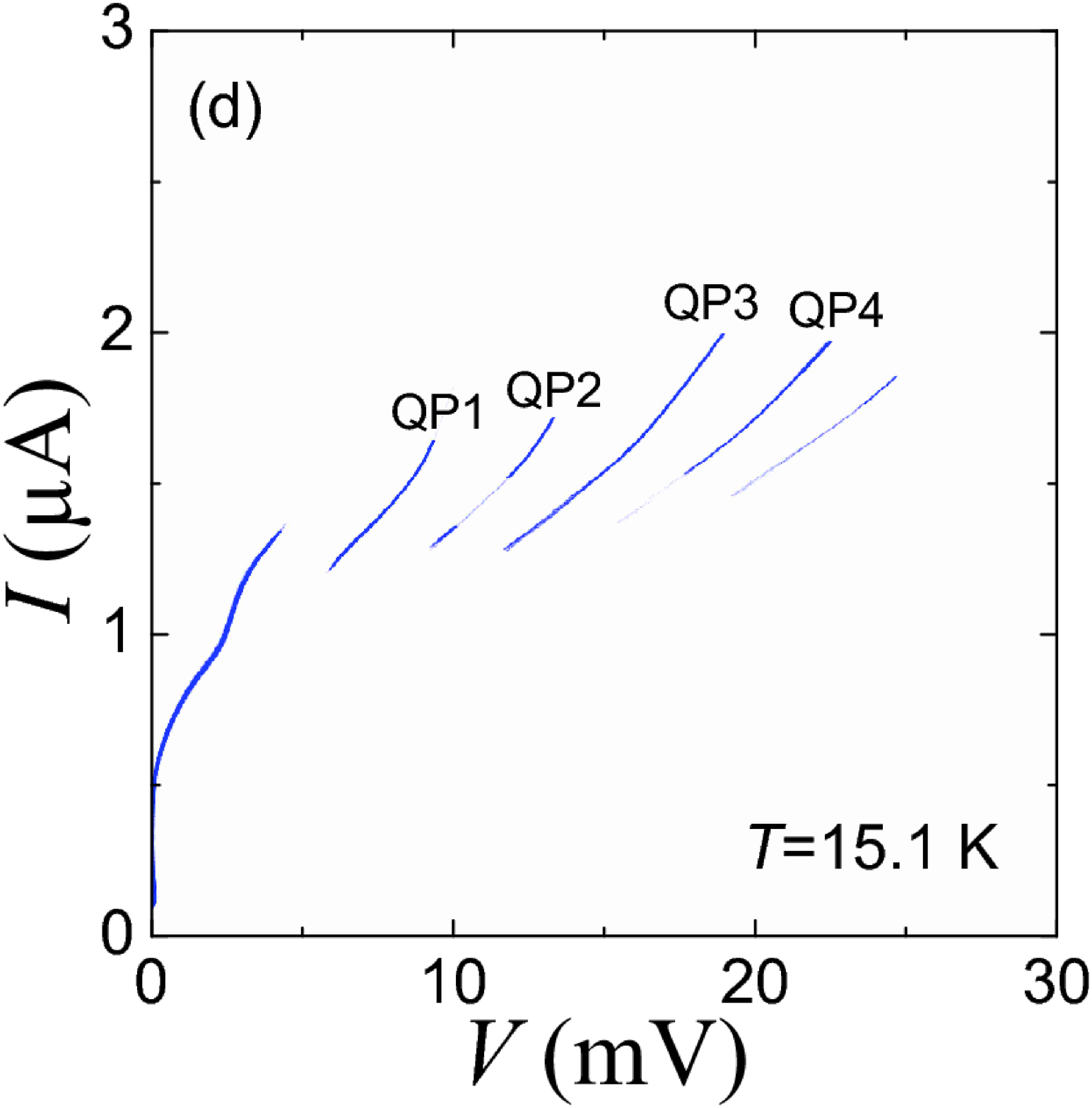}
\end{center}
\caption{\label{fig:1_4T} (Color online) $I$-$V$ curves of the
mesa-1 at $H=1.4$ T and at different $T=2.0~ \text{K}-15.1$ K. At
low $T$, in panels (a) and (b), sequences of hysteretic (high-$Q$)
individual Fiske steps are seen at low bias. At higher bias some
junctions switch into the QP state, but Fiske steps are still
present in the rest of the junctions. The corresponding first four
mixed flux-flow-QP branches are marked (QP1-4). At $T=10.1$ K (c)
these steps smear out and at $T=15.1$ K (d) individual Fiske steps
have vanished, instead a collective, non-hysteretic step is
observed.}
\end{figure}

\section{III. Experimental}

Small mesa structures were fabricated on top of Bi-2212 single
crystals with $T_c=82$ K. Twelve mesas with different sizes were
fabricated simultaneously on every crystal. All of the studied
mesas showed similar behavior. Here we present data for two mesas
on the same slightly underdoped Bi-2212 crystal with areas of
$2.7\times 1.4~ \mu \text{m}^2$ (mesa-1) and $2.0\times 1.7 ~\mu
up\text{m}^2$ (mesa-2). Both mesas contain $N=12$ IJJs. The
results are representative for a large number of mesas made on
crystals with different doping and composition (see Table-I in
Ref.\cite{KatterweFiske}). Details of sample fabrication and of
the experimental set-up can be found in Ref. \cite{KatterweFiske}.

The magnetic field was applied strictly parallel to the
superconducting CuO bilayers, to avoid the intrusion of Abrikosov
vortices.  Eventual entrance of Abrikosov vortices is immediately
obvious in experiment: it causes very strong and irreversible
damping of Fiske resonances and of the Fraunhofer modulation of
the critical current \cite{Katterwe}. Essentially, results
reported here are observable only in the absence of Abrikosov
vortices. Using the rigorous alignment procedure, described in
Ref. \cite{Katterwe}, we were able to prevent Abrikosov vortex
entrance in fields up to 17 T \cite{MR,Polariton}. This is seen
from the field-independence of the $c$-axis QP resistance
\cite{MR} and perfect reversibility of all measured
characteristics \cite{Katterwe,KatterweFiske}.

All measurements are made in the 3-probe configuration. To
simplify data analysis, a contact or a quasiparticle resistance
was subtracted from $I$-$V$ characteristics, as described in Ref.
\cite{Katterwe}. The subtraction is facilitated by the negligible
dependence of the QP resistance on the in-plane magnetic fields
due to the extremely large anisotropy of Bi-2212 (see. e.g. Fig. 3
(d) in Ref. \cite{MR}). To do the subtraction, we first carefully
measured the corresponding branch of the $I$-$V$ at zero magnetic
field. After that we made a high-order polynomial fit of $\ln (I)$
vs. $V$, which is almost linear \cite{KatterwePRL} and can be
fitted with a very high ($\sim \mu$V) accuracy. This fit is then
subtracted from the measured $I$-$V$. When studying
$T$-dependence, this procedure was repeated at each $T$. Such
subtraction simplifies the analysis of Fiske steps, but is not
necessary: Fiske steps can be also measured relative to the
bias-dependent contact or QP voltages.

\section{IV. Results}

Figure \ref{fig:1_4T} shows $I$-$V$ curves (digital oscillograms)
for the mesa-1 at $H=1.4$ T and at different $T$ from 2.0 K to
15.1 K. As the current is increased, the $I$-$V$s switch from the
zero voltage branch to the flux-flow branch, containing sequences
of individual and collective Fiske steps, seen as small
sub-branches in Fig. \ref{fig:1_4T} (a), and ending at the
velocity-matching step. Detailed discussion of the magnetic field
dependence of Fiske and VM steps at low $T$ can be found in Ref.
\cite{KatterweFiske}. Strong hysteresis of Fiske steps indicates
high $Q\gg1$ of the geometric resonances. This is facilitated by
careful alignment of magnetic field, which prevents penetration of
Abrikosov vortices \cite{MR}. With further increase of current
some junctions switch into the QP state, while the rest are
remaining in the flux-flow state. This leads to appearance of
combined QP-FF families of Fiske steps, four of which are
indicated in Fig. \ref{fig:1_4T}(a), (QP1-4) with the number
corresponding to the number of IJJs in the QP state.

The speed can be obtained directly from resonant voltages using
Eqs.(\ref{eq:Vmn}) and (\ref{VMstep}). The corresponding low-$T$
values for several mesas at different Bi-2212 crystals can be
found in Ref. \cite{KatterweFiske}. Fiske steps in Fig.
\ref{fig:1_4T} correspond to slow speed resonances $V_{2,N}=0.27$
mV. At the QP1, QP2 branches another sequence $V_{4,N}=0.54$ mV of
individual Fiske steps is seen. As shown in Refs.
\cite{KatterweFiske,Polariton}, the $V_{VM}$ is proportional to
the field for 2 T $<H <$ 10 T, consistent with Eq. (\ref{VMstep}),
before it gets interrupted by phonon-polariton resonances at
higher fields \cite{Polariton}. In this intermediate field range
the limiting fluxon velocity is close to the out-of-phase velocity
$c_N$.

\begin{figure*}[t]
\begin{center}
   \includegraphics[width=0.9\textwidth]{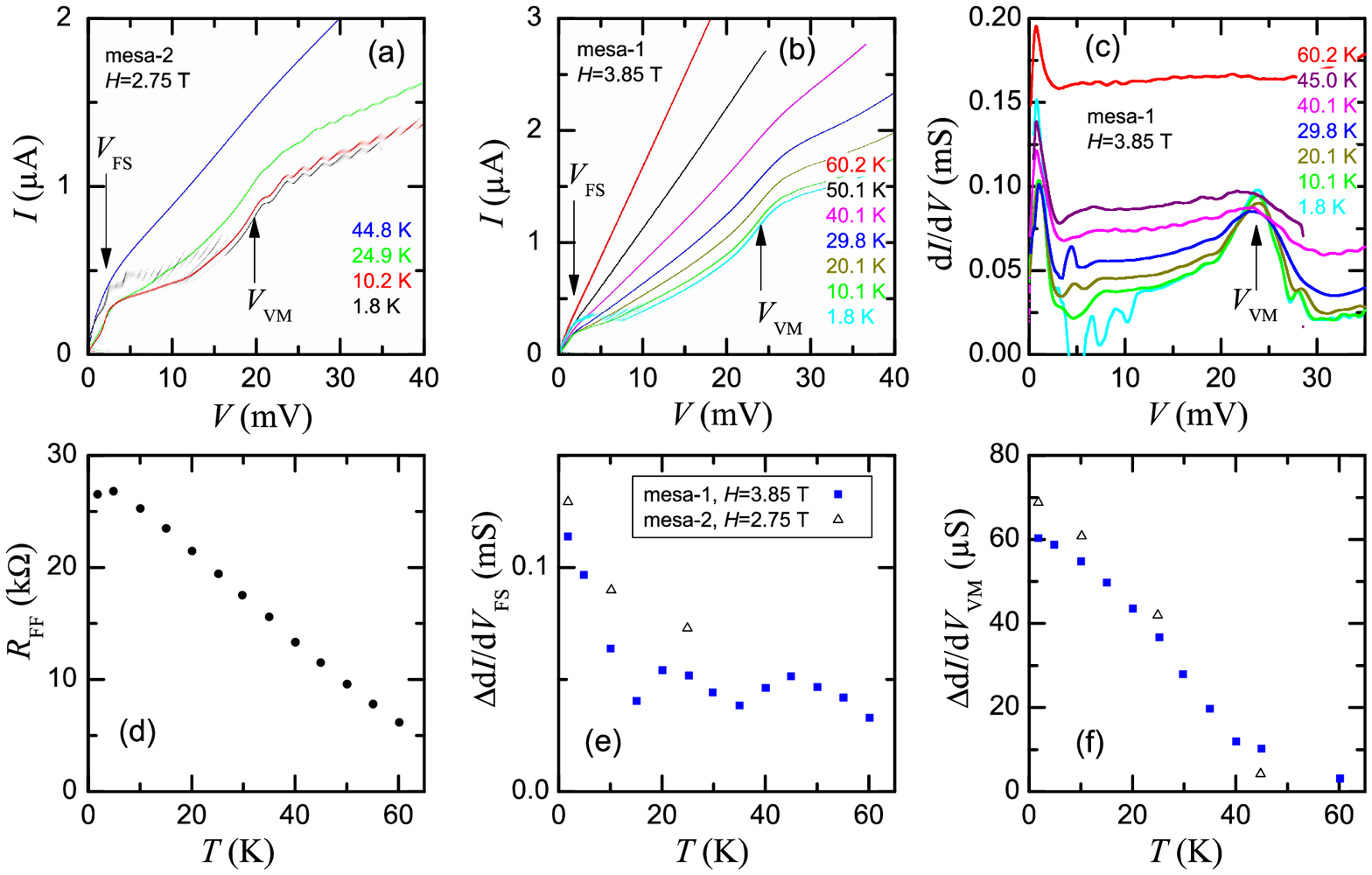}
\end{center}
\caption{\label{fig:4band6b} (Color online) (a) Flux-flow parts of
$I$-$V$ curves of mesa-2 at $H=2.75$ T and at different $T$. A
collective Fiske step (downward arrow) and a velocity matching
step (upward arrow) are seen, followed by QP branches at higher
bias $V>20$ mV. (b) The same for the mesa-1 at $H=3.85$ T. It is
seen that both Fiske and velocity matching steps are rapidly
smeared out with increasing $T$. (c) d$I$/d$V$ curves, numerically
calculated from curves in (b). Distinct peaks correspond to the
collective Fiske and the velocity-matching resonances. (d) $T$-
dependence of the nearly ohmic flux-flow resistance at $V=12$ mV
for the mesa-1 at $H=3.85$ T. (e) and (f) show amplitudes of the
d$I$/d$V$ peaks, corresponding to the collective Fiske step (e)
and the velocity-matching peak (f).}
\end{figure*}

\subsection{A. Temperature dependence of the quality factor}

As seen from Fig. \ref{fig:1_4T}, with increasing temperature, the
amplitude of the individual Fiske steps rapidly decreases. At
$T=10.1$ K steps are smeared out almost completely and at $T=15.1$
K they vanish. This indicates a substantial reduction of the
resonance quality factor. At this temperature only a collective,
non-hysteretic Fiske step is visible at $N\times V_{2,N} \simeq
3.2$ mV, see Fig \ref{fig:1_4T} (d).

Figures \ref{fig:4band6b} (a) and (b) show $I$-$V$s in a wider
$T$-range (a) for the mesa-2 at $H=2.75$ T, and (b) for the mesa-1
at $H=3.85$ T. Collective Fiske steps at $\approx N\times V_{1,N}$
can be seen at low $T$ (indicated by the downward arrows). At
higher bias VM steps are observed (indicated by the upward
arrows). Both mesas show similar behavior: Sharpness of the
collective Fiske and the VM steps rapidly decreases with
increasing temperature. This indicates enhancement of damping,
also seen from reduction of slopes of $I$-$V$ curves with
increasing $T$.

Figure \ref{fig:4band6b} (c) shows d$I$/d$V$ curves, numerically
calculated from the $I$-$V$ curves from (b). Peaks in conductance
correspond to Fiske and VM steps. The decrease of amplitudes of
the steps with increasing $T$ is clearly seen, indicating
reduction of $Q$ at higher temperatures.

According to the sine-Gordon equation, the initial viscous part of
the flux-flow $I$-$V$ should be ohmic with the flux-flow
resistance $R_{FF}$ representing the effective damping
\cite{Scott}. Indeed, from Figs. \ref{fig:4band6b} (a) and (b) it
is seen that the flux-flow $I$-$V$ is nearly ohmic at $10 < V <
20$ mV. This allows accurate evaluation of the bare (non-resonant)
$R_{FF}(T)$. It is shown in panel (d) for $V=12$ mV ($\sim $ 1 mV
per junction). The $T$-dependence of $R_{FF}$ is almost identical
to the low-bias $c$-axis QP resistance $R_{QP}(T, H=0)$
\cite{KatterwePRL}, proving that the $R_{FF}(T)$ dependence is
predominantly determined by ``freezing out" of quasiparticles. At
low $T$ and moderately low $H$ the value of $R_{FF}$ is slightly
lower than $R_{QP}$, which may indicate presence of additional
damping mechanisms, such as the in-plane QP damping
\cite{KoshelevInplane}, or generation of phonons via
electrostriction \cite{Polariton}. At higher $H$, $R_{FF}=R_{QP}$
(see e.g. Fig. 3 (d) from Ref.\cite{MR}).

Figures \ref{fig:4band6b} (e) and (f) represent $T$-dependencies
of bare amplitudes of conductance peaks at the collective Fiske
step and the VM step, respectively. The peak amplitudes were
obtained by subtracting the background flux-flow conductance
$R_{FF}^{-1}$. It is seen that resonances in both mesas exhibit
similar $T$-dependencies: At low $T$, peaks are high, i.e.,
quality factors of resonances are large $Q\gg 1$, but they start
to rapidly decrease with increasing $T$. Comparison with the
effective flux-flow resistance $R_{FF}$, shown in panel (d),
indicates that the scale for variation of peak amplitudes is
similar to $R_{FF}(T)$. Therefore, both resonances roughly follow
Eq. (\ref{eq:Q}) with $R \simeq R_{FF}(T)$.

\begin{figure}[t]
\includegraphics[width=0.4\textwidth]{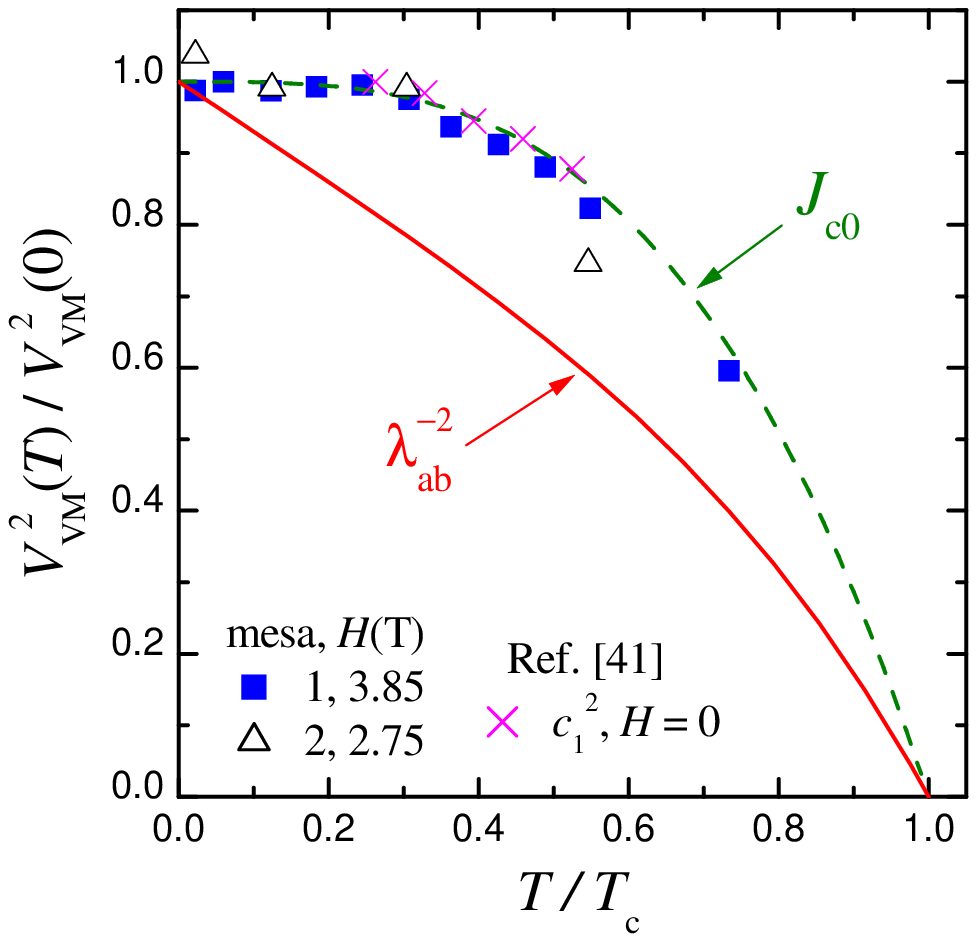}
\caption{\label{fig3} (Color online) Normalized temperature
dependence of the square of the velocity-matching voltage
$V_{VM}\propto c_N$. Solid line represents typical $T$-dependence
of $\lambda_{ab}^{-2}$ from Ref. \cite{Lambda,Lambda2}. Dashed
line represents the $T$-dependence of the fluctuation free
Josephson critical current density $J_{c0} \propto \omega_p^2$
from Ref. \cite{Collapse}.}
\end{figure}

\subsection{B. Temperature dependence of the Swihart velocity}

Both Fiske and VM steps in the considered case correspond to
propagation of waves, respectively fluxons, with the velocity
$\simeq 3.2 \times 10^5$ m/s \cite{KatterweFiske} is close to the
expected value of the slowest out-of-phase velocity $c_N$, Eq.
(\ref{cN}). It is almost 1000 times slower than $c$, not because
of extraordinary large dielectric constant, but because of
extraordinary large kinetic inductance of atomically thin
superconducting layers in Bi-2212, see Eq. (2). According to Eq.
(\ref{cN}), $c_N(T)$ should depend solely on $1/\lambda_{ab}(T)$.
Thus voltages of Fiske and VM steps should provide a direct
information on absolute values of $1/\lambda_{ab}(T)$.

Squares and triangles in Fig. \ref{fig3} represent measured
$T$-dependencies of $V_{VM}^2$ for both studied mesas. Crosses in
Fig. \ref{fig3} represent fast geometrical resonance voltages,
reported recently by Benseman and co-workers on large Bi-2212
mesas at zero field \cite{Tim}. Apparently, our data for the
slowest resonances coincide with their data for the fast resonance
within the measured $T$-range.

Lines in Fig. \ref{fig3} represent typical temperature
dependencies of $\lambda_{ab}^{-2}$ for cuprates
\cite{Lambda,Lambda2} and the fluctuation-free $c$-axis critical
current density $J_{c0}$ for Bi-2212 IJJs
\cite{Kleiner92,Collapse}. The latter is similar to
$\omega_p^{2}(T)$, measured by the Josephson plasma resonance
\cite{Plasma} and to $\lambda_c^{-2}(T)$ obtained from surface
impedance measurements \cite{Lambda_c}, consistent with
Eqs.(\ref{OmegaP},\ref{Lambda_c}). It is seen that
$\lambda_{ab}^{-2}$ and $J_{c0}$ exhibit distinctly different
behavior at low $T$: $J_{c0}(T)$ is flat, while
$\lambda_{ab}^{-2}(T)$ has a linear $T$-dependence due to the
d-wave symmetry of the order parameter \cite{Lambda,Lambda2}.
Clearly, experimental $V_{VM}^2(T)$ follow $J_{c0}(T)$ rather than
the expected $\lambda_{ab}(T)^{-2}$ dependence.

\section{V. Discussion}

At low $T$, the obtained speed of EM waves $\simeq 3.2 \times
10^5$ m/s agrees with the expected out-of-phase mode velocity
$c_N$, Eq. (\ref{cN}) for reasonable parameters $t/\varepsilon_r
=0.1$ nm and $\lambda_{ab}(T=0) \simeq 200$ nm
\cite{Lambda,Lambda2,Lambda_abc,Trunin}. Thus the ICJ model does
provide a correct value of the Swihart velocity at low $T$. It
also provides correct $T$-dependencies of the Josephson plasma
frequency \cite{Plasma} and $\lambda_c$ \cite{Lambda_c},
$\omega_p(T)\propto \lambda_c^{-1}(T)\propto \sqrt{J_{c0}(T)}$,
see Eqs.(\ref{OmegaP},\ref{Lambda_c}). Therefore, it is surprising
that the $T$-dependence of the effective penetration depth deduced
from resonant voltages is different from $\lambda_{ab}(T)$,
obtained from surface impedance measurements
\cite{Lambda,Lambda2,Lambda_abc,Trunin}. Below we mention several
possible reasons for such a discrepancy.

\subsection{A. Possible origin of discrepancy with surface impedance measurements}

Derivation of the ICJ model is based on the assumption that field
and current distributions within each superconducting layer can be
described by the local 2nd London equation \cite{SBP}. However,
this assumption most likely breaks down in atomic scale IJJs (see
the condition (4.3) in Ref. \cite{Economou}).

To understand the reported discrepancy it is, first of all,
necessary to understand the difference in local current and field
distributions. In surface impedance measurements, the external
electromagnetic field is screened at the depth $\lambda_{ab} \sim
200$ nm from the surface of the superconductor. This induces
similar (in-phase) screening currents in a fairly large number $N
\sim 130$ of IJJs. To the contrary, at the out-of-phase
geometrical resonances the current varies at the atomic scale, as
shown in Fig. \ref{fig1} (b).

{\it i. Non-locality of supercurrent}

The most obvious question is to what extent Cooper pairs are
localized in every CuO bi-layer.
The very existence of the $c$-axis critical current indicates that
the localization is incomplete. This
can be particularly significant for the out-of-phase mode, when
Cooper pairs are forced to move in opposite directions in neighbor
layers, see Fig. \ref{fig1} (b). Qualitatively such delocalization
will lead to larger effective penetration depth.

{\it ii. Nonlocal Josephson electrodynamics}

Another type of non-locality in thin layer junctions was
considered in Ref. \cite{Mints}. With decreasing $d$, the
effective screening length $\Lambda/2$ Eq.(3) increases and
approaches the Pearl length $\lambda_P=\lambda_S^2/d$. To the
contrary, the Josephson penetration depth $\lambda_J$ decreases
$\propto \Lambda^{-1/2}$ see Eq.(\ref{Lambda_J}). For IJJs
$\lambda_J < 1~\mu$m \cite{Katterwe} is much smaller than
$\lambda_P \sim 10~\mu$m even at $T=0$. Such a mismatch changes
the dispersion relation of electromagnetic waves \cite{Mints}.

{\it iii. Retardation effects}

Retardation effects appear in transmission lines when the time
(phase velocity) required to transfer charge within a layer is
comparable or faster than that for electromagnetic waves outside
the layer \cite{Economou}. Specific for IJJs is that the
out-of-phase electromagnetic wave velocity is so slow $\sim 10^5
~$ m/s, that it becomes comparable to the electronic Fermi
velocity. This may affect the dispersion relation.

{\it iv. Frequency dependence}

The effective penetration depth in superconductors depends not
only on $T$ but also on frequency $\lambda(T,\omega)$. It
originates from a significant $(T,\omega)$ dependence of complex
conductivity in a superconductor \cite{Ngai}. The most obvious
difference between static and high-frequency $\lambda(T)$ is that
the latter does not diverge at $T\rightarrow T_c$, but approaches
the finite normal skin-depth. This may flatten-out $T$-dependence
of high frequency Fiske steps, compared to static $1/\lambda(T)$
\cite{Ngai}.

Surface impedance measurements are typically performed at $\sim
10$ GHz frequency. In comparison, the studied Fiske and VM step
voltages are $\sim 1$ mV per junction, see Fig. \ref{fig:4band6b}.
According to the ac-Josephson relation, this corresponds to $\sim
500$ GHz. The significant difference in frequencies may lead to a
significant difference in the effective $\lambda$.

At even higher THz frequencies, the frequency dependence of the
dielectric function $\varepsilon_r(\omega)$ in isolating BiO
layers becomes significant. As shown in Ref. \cite{Polariton}, the
speed of electromagnetic waves slows down dramatically, when the
frequency approaches the transverse optical phonon frequencies.

{\it v. Non-linear effects}

Eq.(\ref{c_n}) was derived by linearization of the coupled
sine-Gordon equation and is valid for small amplitude EM waves $a
\ll 1$. However, at high quality geometrical resonances the
amplitude may be large $a\sim 2\pi$ and non-linearity of the
sine-Gordon equation may affect the dispersion relation.

The penetration depth depends on the absolute value of the current
density. Close to the depairing current density, $\lambda$ rapidly
increases. At geometrical resonances the amplitude of the in-plane
current density is given by Eq. (14). It depends on the amplitude
$a$, which can be $\sim 1$ for $Q\gg 1$. An estimation for $a=1$
and $L/m = 1 ~\mu$m yields $J_{ac}\sim 10^6$ A/cm$^2$, comparable
to the maximum in-plane current density \cite{Yurgens}.

Both types of non-linear effects increase with increasing the
quality factor of resonances. Since $Q\gg 1$ only at low $T$,
nonlinear corrections can be significant at low $T$, but less so
at elevated temperatures.

\subsection{B. Implications for coherent Josephson oscillators}

As mentioned in the Introduction, stacked IJJs are considered as
possible candidates for high power THz oscillators
\cite{Ozyuzer,Wang,Nonequilibrium,Hu,Klemm,Breather,Tim}. A large
energy gap in Bi-2212 \cite{KatterwePRL,MR} allows generation of
electromagnetic radiation with frequencies in excess of $10$ THz.
For example, recently polariton generation with frequencies up to
$\sim 13$ THz was reported \cite{Polariton}. Moreover, strong
electromagnetic coupling of IJJs facilitates phase-locking of many
junctions, which may lead to coherent amplification of the
emission power \cite{Note1}.

Realization of a flux-flow oscillator \cite{Koshelets}, based on
fluxon motion in the in-plane magnetic field
\cite{Cherenkov,Bae2007,FFlowHatano,FiskeInd}, encounter a
difficulty, associated with instability of the rectangular fluxon
lattice. It can be stabilized by geometrical confinement in small
mesas \cite{Katterwe} or by interaction with infrared optical
phonons \cite{Polariton}. But usually fluxon-fluxon repulsion
promotes the triangular fluxon lattice, corresponding to the
out-of-phase state, which leads to destructive interference and
negligible emission \cite{TheoryFiske}.

High quality geometrical resonances improve operation of a stacked
oscillator is several ways: (i) they amplify the emission power
$\propto Q^2$ \cite{TheoryFiske}; (ii) they narrow the radiation
linewidth $\propto 1/Q$ \cite{TheoryFiske}; (iii) they can {\it
force} phase-locking of junctions. Numerical simulations have
demonstrated that large amplitude standing waves, $a\sim 1$, can
superimpose their symmetry on the fluxon lattice
\cite{KatterweFiske}. Such a non-linear synchronization requires
high $Q$ because $a \propto 1/Q$.

The reported rapid decrease of the quality factor with increasing
temperature indicates that self-heating is detrimental for the
coherent Josephson oscillator and ultimately limits the emission
power from large Bi-2212 mesas \cite{Breather}. On the other hand,
$T$-dependence of the Swihart velocity facilitates fairly
broad-range tuning of the resonance frequency, as seen from Fig.
\ref{fig3}. This may be beneficial for the oscillator \cite{Tim}.

\section{Conclusions}
To conclude, we have studied $T$-dependence of geometrical and
velocity matching resonances in small Bi-2212 mesa structures. We
reported strong $T$-dependence of the quality factors, which is
large at low $T$, but rapidly decreases with increasing $T$,
already at $T>10$ K. Above $T \sim 60$ K $\sim 0.8 T_c$ resonances
are almost fully damped. This observation is consistent with
previous observations of strongly underdamped phase dynamics at
low $T$ \cite{Collapse}, leading to relatively high macroscopic
quantum tunnelling temperature in IJJs \cite{Lombardi,MQT,MQT2},
and with the reported collapse into overdamped dynamics at $T/T_c
\sim 0.8$ \cite{Collapse}. The rapid decrease of $Q(T)$ indicates,
that self-heating is detrimental for operation of the coherent THz
oscillator and ultimately limits its performance \cite{Breather}.
On the other hand, $T$-dependence of the Swihart velocity
facilitates a broad-range tuning of the resonance frequency, which
may be beneficial for the oscillator.

Our analysis of $T$-dependence of resonant voltages revealed that
the effective penetration depth, that determines the kinetic
inductance and the speed of electromagnetic waves in intrinsic
Josephson junctions, see Eq. (\ref{Lsquare}), is exhibiting a flat
$T$-dependence at low, resembling $T$-dependence of the $c$-axis
critical current. It is distinctly different from the linear
$T$-dependence of $\lambda_{ab}(T)$, obtained from surface
impedance measurements \cite{Lambda,Lambda2}. We argued that
non-trivial physical phenomena, such as break-down of the local
London approximation at the atomic scale, are responsible for this
distinct discrepancy, which deserves further theoretical
consideration.

Acknowledgements: We are grateful to A. Rydh and H. Motzkau for
assistance in experiment and to the Swedish Research Council and
the SU-Core Facility in Nanotechnology for financial and technical
support, respectively.

\end {document}